\documentclass[iop]{emulateapj}
  \usepackage{timesfonts}

\shorttitle{MAGNETIC ACTIVITY CYCLES IN $\epsilon$~ERI}
\shortauthors{METCALFE ET AL.}

\begin{document}

\title{Magnetic Activity Cycles in the Exoplanet Host Star $\epsilon$~Eridani}

\author{T.~S.~Metcalfe\altaffilmark{1}, 
	A.~P.~Buccino\altaffilmark{2,3},  
	B.~P.~Brown\altaffilmark{4},
	S.~Mathur\altaffilmark{1,5}, 
	D.~R.~Soderblom\altaffilmark{6}, 
 	T.~J.~Henry\altaffilmark{7}, \\
	P.~J.~D.~Mauas\altaffilmark{2,3},
	R.~Petrucci\altaffilmark{2,8},
        J.~C.~Hall\altaffilmark{9},
	S.~Basu\altaffilmark{10}}

\altaffiltext{1}{Space Science Institute, 4750 Walnut St.\ Suite 205, Boulder CO 80301 USA}
\altaffiltext{2}{Instituto de Astronom{\'\i}a y F{\'\i}sica del Espacio (CONICET), C.C.\ 67 Sucursal 28, C1428EHA-Buenos Aires, Argentina}
\altaffiltext{3}{Departamento de F\'\i sica, FCEyN, Universidad de Buenos Aires, 1428, CABA, Argentina}
\altaffiltext{4}{Department of Astronomy \& Center for Magnetic Self-Organization, University of Wisconsin, Madison WI 53706-1582 USA}
\altaffiltext{5}{High Altitude Observatory, NCAR, P.O.\ Box 3000, Boulder CO 80307 USA}
\altaffiltext{6}{Space Telescope Science Institute, 3700 San Martin Dr., Baltimore MD 21218 USA}
\altaffiltext{7}{Department of Physics and Astronomy, Georgia State University, Atlanta GA 30302 USA}
\altaffiltext{8}{Visiting Astronomer, Complejo Astron{\'o}mico El Leoncito (CASLEO).}
\altaffiltext{9}{Lowell Observatory, 1400 West Mars Hill Road, Flagstaff AZ 86001 USA}
\altaffiltext{10}{Department of Astronomy, Yale University, P.O.\ Box 208101, New Haven CT 06520 USA}

  \submitted{The Astrophysical Journal Letters, (ACCEPTED)}

\begin{abstract}

The active K2 dwarf $\epsilon$~Eri has been extensively characterized, 
both as a young solar analog and more recently as an exoplanet host star. 
As one of the nearest and brightest stars in the sky, it provides an 
unparalleled opportunity to constrain stellar dynamo theory beyond the 
Sun. We confirm and document the 3~year magnetic activity cycle in 
$\epsilon$~Eri originally reported by Hatzes and coworkers, and we examine 
the archival data from previous observations spanning 45 years. The data 
show coexisting 3~year and 13~year periods leading into a broad activity 
minimum that resembles a Maunder minimum-like state, followed by the 
resurgence of a coherent 3~year cycle. The nearly continuous activity 
record suggests the simultaneous operation of two stellar dynamos with 
cycle periods of $2.95\pm0.03$ years and $12.7\pm0.3$ years, which by 
analogy with the solar case suggests a revised identification of the 
dynamo mechanisms that are responsible for the so-called ``active'' and 
``inactive'' sequences as proposed by B\"ohm-Vitense. Finally, based on 
the observed properties of $\epsilon$~Eri we argue that the rotational 
history of the Sun is what makes it an outlier in the context of magnetic 
cycles observed in other stars (as also suggested by its Li depletion), 
and that a Jovian-mass companion cannot be the universal explanation for 
the solar peculiarities.

\end{abstract}

\keywords{stars: activity---stars: chromospheres---stars: 
individual(HD~22049)---surveys}

\section{BACKGROUND}\label{SEC1}

The study of stellar magnetic activity cycles dates back to the 1960's, 
when Olin Wilson began monitoring Ca~{\sc ii} H and K (Ca~HK) emission for 
a sample of stars from the Mount Wilson Observatory (MWO) to provide some 
context for our understanding of the 11~year sunspot cycle \citep{wil78}. 
The Mount Wilson survey continued for several decades \citep[see][and 
references therein]{bal95}, and ultimately documented activity cycles and 
rotation periods for dozens of stars \citep{sb99}. These observations 
revealed two distinct relationships between the activity cycle period and 
the rotation period, with an active ``A'' sequence including stars 
rotating more than 300 times for each activity cycle, and an inactive 
``I'' sequence with stars rotating fewer than 100 times per activity 
cycle. This pattern led \cite{bv07} to suggest that there may be two 
different dynamos operating inside the stars, with the active sequence 
representing a dynamo driven by rotational shear in the near-surface 
layers, and the inactive sequence driven by a so-called {\it tachocline} 
at the base of the outer convection zone. Some stars in the Mount Wilson 
sample exhibit two distinct cycle periods, suggesting that the two dynamos 
can operate simultaneously.

The K2V star $\epsilon$~Eridani ($\epsilon$~Eri $\equiv$ HD~22049, V=3.7, 
B$-$V=0.88) is a young solar analog with a stellar activity record that 
stretches back to 1968. The first 24 years of observations were published 
in \cite{gb95}, who determined a rotation period of 11.1~days and found 
evidence of a 5~year activity cycle. This cycle period was confirmed by 
\cite{bm08} from a joint analysis of the Mg~{\sc ii} h and k lines in 
archival IUE spectra and more recent Ca~HK observations from the Complejo 
Astron{\'o}mico El Leoncito (CASLEO) in Argentina. Additional Ca~HK data 
were obtained by \cite{hal07} with the Solar-Stellar Spectrograph at 
Lowell Observatory, though with less regularity because the star is 
slightly cooler than the close solar analogs that dominate their sample. 
As a bright nearby star \cite[$d=3.2$~pc;][]{vl07}, $\epsilon$~Eri also 
has a precisely measured radius from interferometry \citep[$R=0.74\pm0.01 
R_\odot$;][]{ba12}. A further constraint for stellar dynamo modeling comes 
from a measurement of surface differential rotation, derived from 35~days 
of photometry by the {\it MOST} satellite \citep{cro06}.

  \begin{figure*}[t] 
  \centerline{\includegraphics[angle=270,width=5.5in]{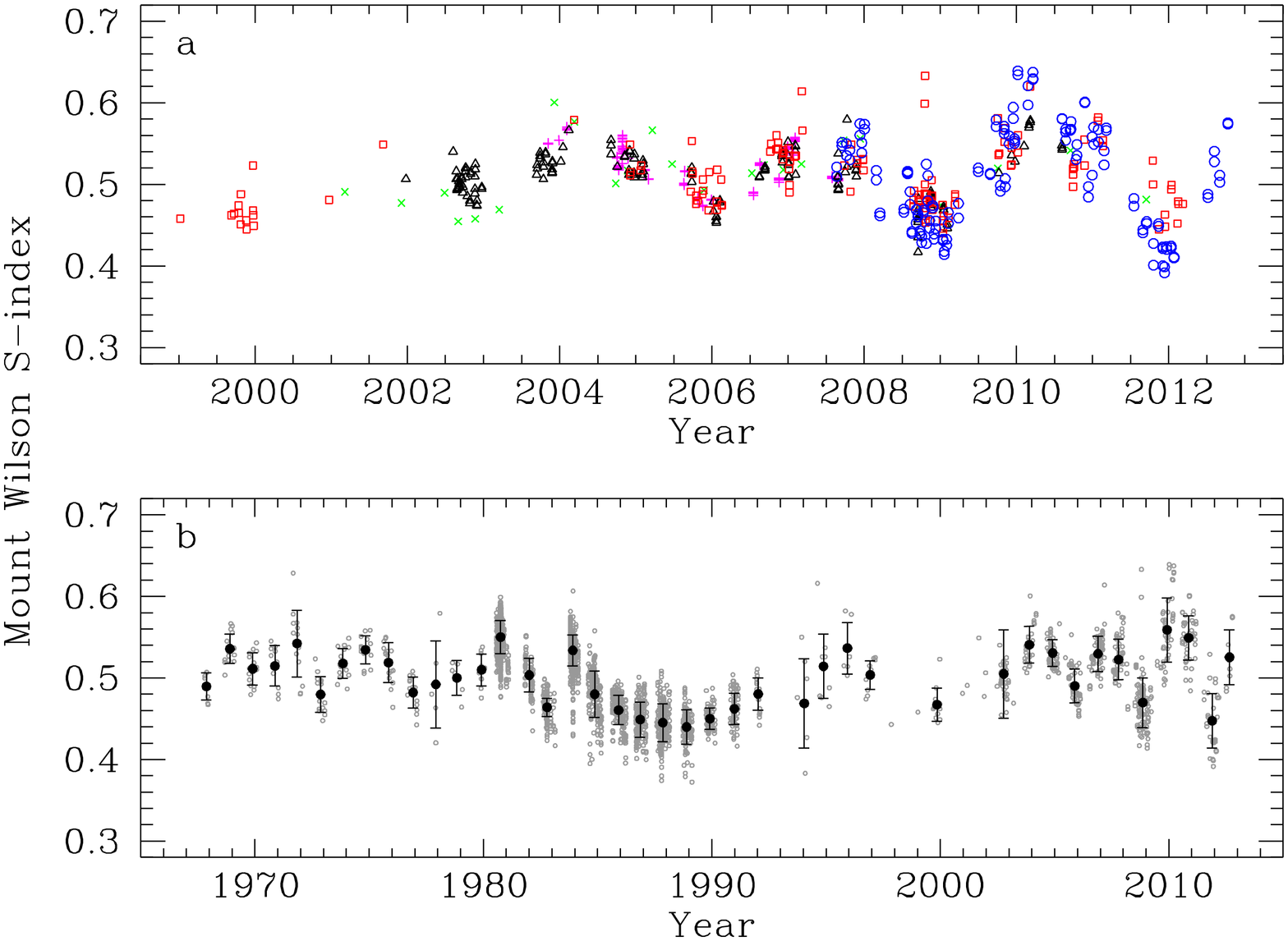}} 
  \caption{Chromospheric activity measurements of the K2V star 
  $\epsilon$~Eri. {\bf a:} Recent data from SMARTS ($\circ$), Lowell 
  Observatory ($\Box$), and CASLEO ($\times$), along with previously 
  published measurements from CPS \citep[$\bigtriangleup$,][]{if10} and 
  HARPS \citep[$+$,][]{ab12}. {\bf b:} Archival data from Mount Wilson 
  \cite[1968-1992;][]{gb95} and the more recent observations (grey points) 
  with seasonal means ($\bullet$) and uncertainties reflecting the 
  standard deviation within each season.\label{fig1}} 
  \end{figure*} 

Observations of the variable radial velocity (RV) of $\epsilon$~Eri were 
first reported by \cite{cam88}. Early claims by \cite{wal95} of periodic 
variations were eventually corroborated by \cite{cum99}, who identified a 
period near 7 years. \cite{hat00} used additional RV data to confirm the 
$\sim$7 year period, interpreting it as the reflex motion from an 
eccentric Jovian-mass exoplanet; they ruled out stellar activity as the 
source of the variations from an analysis of Mount Wilson data between 
1980-1999. The Ca~HK observations revealed periodic signals near 20 years 
(the length of the data set) and 3 years, but nothing at the orbital 
period of the presumed planet. The planetary nature of the companion was 
later confirmed by \cite{ben06} who measured the astrometric orbit using 
the {\it Hubble Space Telescope}, yielding an inclination angle 
$i=30\fdg1\pm3\fdg2$ \citep[consistent with that of an observed dust 
ring;][]{gre05} and determining the absolute mass of the planet, $M=1.5\ 
M_J$. More recent RV measurements from the HARPS spectrograph imply that 
the properties of the planet may need to be revised, but still support a 
period near 7 years \citep{ab12}.

We present new observations of the magnetic activity variations in 
$\epsilon$~Eri from synoptic Ca~HK measurements obtained since 2007 with 
the Small and Moderate Aperture Research Telescope System (SMARTS) 
\mbox{1.5-m} telescope at Cerro Tololo Interamerican Observatory (CTIO). 
We provide an overview of our data as well as corroborating measurements 
from other surveys in section~\ref{SEC2}, and in section~\ref{SEC3} we 
evaluate the dominant periodicities over time using archival observations 
spanning 45 years. We conclude in section~\ref{SEC4} with a discussion of 
the implications of these results for stellar dynamo modeling and future 
observations.

\section{OBSERVATIONS \& ARCHIVAL DATA}\label{SEC2}

The SMARTS southern HK project \citep{met09,met10} began in August 2007 
with the primary objective of characterizing magnetic activity cycles for 
the brightest stars (V$<$6) in the southern hemisphere. The 58 solar-type 
stars in the sample, defined as a subset of the \cite{hen96} sample, 
included all of the most likely future asteroseismic targets of the 
Stellar Observations Network Group \citep[SONG;][]{gru08}. Several targets 
near the celestial equator, including $\epsilon$~Eri, provided an overlap 
with the Mount Wilson and Lowell surveys, allowing the calibration of 
derived Ca~HK \mbox{S-indices} onto the Mount Wilson system.

Since August 2007, we have used the {\it RC Spec} instrument on the SMARTS 
\mbox{1.5-m} telescope to obtain 141 low-resolution spectra ($R\sim2500$) 
of $\epsilon$~Eri on 69 distinct epochs. Standard IRAF\footnote{IRAF is 
distributed by the National Optical Astronomy Observatory, which is 
operated by the Association of Universities for Research in Astronomy 
(AURA) under cooperative agreement with the National Science Foundation.} 
routines were used on the 60\,s integrations to apply bias and flat field 
corrections, and a wavelength calibration was applied using a reference 
\mbox{He-Ar} spectrum obtained just prior to each pair of stellar 
exposures. Following \cite{dun91}, we integrated the calibrated spectra in 
1.09~\AA\ triangular bandpasses centered on the cores of the Ca H 
(396.8~nm) and K (393.4~nm) lines and compared them to 20~\AA\ continuum 
regions from the wings of the lines to generate a CTIO chromospheric 
activity index, $S_{\rm CTIO}$. We used data for 26 targets that were 
observed contemporaneously with the Solar-Stellar Spectrograph at Lowell 
Observatory to make the conversion to Mount Wilson indices ($S_{\rm 
MWO}$). Our SMARTS observations of $\epsilon$~Eri are listed in 
Table~\ref{tab1}, where the quoted uncertainties represent the internal 
errors and do not include the systematic uncertainty ($\sigma_{\rm 
sys}\sim +0.007$) from the conversion between the CTIO and MWO indices.

We used additional observations of $\epsilon$~Eri from previously 
published surveys to corroborate our recent SMARTS measurements and to 
extend the time baseline of our analysis. The MWO data from 1968--1992 
were published by \cite{gb95}. Although observations were collected at 
Mount Wilson for nearly an additional decade beyond 1992, those data have 
never been released. Fortunately, the data from Lowell Observatory begin 
in 1994 and continue to the present \citep[see][]{hal07}. As the Lowell 
series is the longest among the recent measurements, we used these 
observations to rescale the other data to the MWO system. The southern 
survey at CASLEO includes measurements of $\epsilon$~Eri from 2001--2011 
\citep{mau12}. Additional published data between 2002--2010 are available 
from the California Planet Search \citep[CPS;][]{if10}, and between 
2004--2008 from the HARPS spectrograph \citep{ab12}. In each case we 
placed the \mbox{S-index} measurements on the Mount Wilson system from 
their overlap with the Lowell data, yielding multiplicative scale factors 
of 1.04 (CASLEO), 1.09 (CPS), and 1.32 (HARPS). The results presented in 
section~\ref{SEC3} are insensitive to small changes in these factors. From 
the combined data we calculated seasonal means for each year that had more 
than three observations, including uncertainties that reflect the standard 
deviation within each season. The observed scatter is dominated by actual 
variations in the \mbox{S-index} caused by rotational modulation of 
individual active regions, but it also contains a small contribution from 
the statistical uncertainties of the individual measurements (cf.\ 
Table~\ref{tab1}).

The chromospheric activity measurements of $\epsilon$~Eri are shown in 
Figure~\ref{fig1}. The top panel includes the recent measurements from 
SMARTS ($\circ$, see Table~\ref{tab1}), as well as the data from Lowell 
($\Box$), CASLEO ($\times$), CPS ($\bigtriangleup$), and HARPS ($+$). The 
bottom panel includes all of the recent data as well as the archival 
measurements from Mount Wilson (grey points) along with our calculated 
seasonal means and those from \cite{gb95}.

\section{RESULTS \& INTERPRETATION}\label{SEC3}

\begin{figure}[t]
  \centerline{\includegraphics[angle=0,width=3.3in]{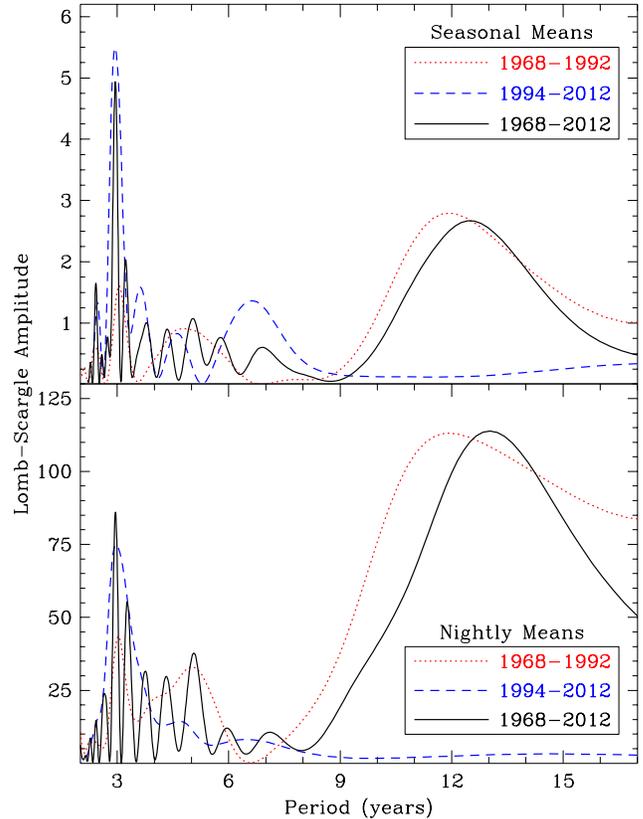}}
  \caption{Lomb-Scargle periodograms of the seasonal (top panel) and nightly 
  (bottom panel) mean \mbox{S-index} measurements from the MWO data 
  1968-1992 (dotted), the recent data 1994-2012 (dashed) and the full data 
  set 1968-2012 (solid). Significant variations are detected in the complete 
  time series with periods near 3~years and 13~years, corresponding to 
  cycles on the inactive and active sequence respectively for a star with a 
  rotation period near 11~days \citep{bv07}.\label{fig2}}
  \end{figure}

We determined the significant periodicities in our chromospheric activity 
measurements by passing both the seasonal and nightly mean 
\mbox{S-indices} through a Lomb-Scargle periodogram \citep{lom76,sca82}. 
In each case, we also performed the analysis on two subsets of the data, 
between 1968--1992 from MWO and on the more recent measurements between 
1994--2012, to understand the dominant source of various periodicities. 
The results are shown in Figure~\ref{fig2}, where the top and bottom 
panels show the periodogram of the seasonal and nightly means, 
respectively. The early data from MWO (dotted lines) reveals a primary 
periodicity near 13 years and smaller peaks near 3~years and 5~years. The 
13~year period appears to be rooted in the broad activity minimum between 
1985--1992 in the MWO data, but it is also evident in the declining 
activity level of successive minima since 2006 (see Figure~\ref{fig1}). 
The more recent data (dashed lines) are dominated by a period near 3 
years, but they also show a weak signal near 7~years which corresponds to 
the orbital period of the exoplanet. This 7~year period is present in the 
individual observations of the longer time series measurements from both 
Lowell and CPS, while the 3~year signal is present in each of the recent 
data sets individually. The longest time series from Lowell resolves the 
power between 3--5 years into multiple components, among which the 3~year 
signal is the strongest. This is also reflected in the periodogram of the 
full data set (solid lines), which shows significant periodicities at 
$2.95\pm0.03$ years and $12.7\pm0.3$ years (both with false alarm 
probability $<10^{-6}$ from the nightly means) along with several weaker 
peaks between 3--7 years. Simulations of the two dominant periods reveal 
all of these peaks to be artifacts of the time sampling. An additional 
peak at longer periods of 20--35 years is present in both periodograms, 
but we exclude it from our analysis because it is correlated with the 
length of the adopted data set.

\begin{figure}[t]
  \centerline{\includegraphics[angle=90,width=3.3in]{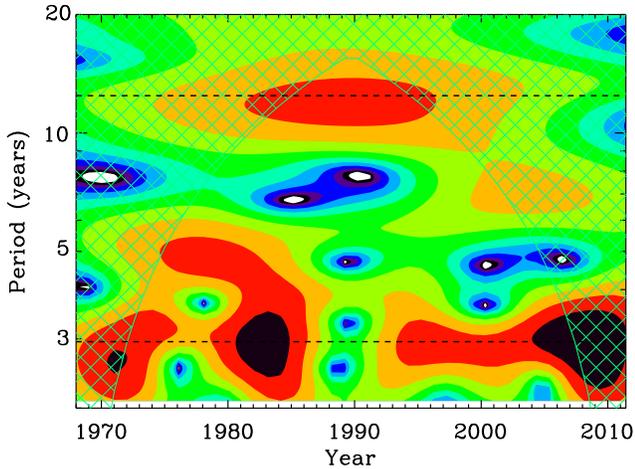}}
  \caption{Wavelet spectrum of the seasonal mean \mbox{S-index} measurements 
  from 1968-2012, showing the relative strength of the cycle periods over 
  time. The hatched region marks the area outside of the cone of influence 
  where signals can be reliably measured with the method, while the color 
  scale indicates the significance of the signal from the weakest (white and 
  blue) to the strongest (black and red).\label{fig3}}
  \end{figure}

To investigate the strength of these periodic signals over time, we 
performed a wavelet analysis of the seasonal mean \mbox{S-index} 
measurements \citep{tc98}. Such an analysis essentially calculates a 
periodogram for overlapping subsets of the time series, where the length 
of the subset must be proportional to the periodicity under investigation. 
Consequently, there are regions of the wavelet spectrum (outside of the 
so-called ``cone of influence'') that suffer from edge effects, where the 
signal is attenuated. The results are shown in Figure~\ref{fig3}, where 
the border of the cone of influence is indicated with a hatched region and 
the significance of the signal is shown with a color scale going from the 
weakest (white and blue) to the strongest (black and red). The 2.95~year 
and 12.7~year periods are indicated with dashed horizontal lines. The 
2.95~year period maintains its strength through most of the duration of 
the time series, with the exception of the late 1980's to early 1990's 
through the broad activity minimum. During this interval, the signal is 
dominated by the 12.7~year periodicity, which remains strong inside the 
cone of influence. The spurious 5~year signal in the MWO data appears only 
in the early observations, while the 7~year artifact in the recent data is 
also transient and even less prominent. Due to the limited time-resolution 
of the wavelet analysis for longer periods, it is difficult to determine 
whether the 2.95~year and 12.7~year periods coexist simultaneously, or if 
they alternate instead. If the 2.95~year period actually disappeared 
during the broad activity minimum, it may represent the first observation 
of another star entering (and later emerging from) a Maunder minimum-like 
state for the short cycle.

It is striking that $\epsilon$~Eri displays such short magnetic activity 
cycles. Many models of the solar dynamo favor a flux-transport paradigm, 
and typically the slow meridional circulations set the cycle timescale for 
a dynamo operating in the tachocline at the base of the outer convection 
zone \citep[e.g.,][]{dg06}. From three-dimensional (3-D) models of stellar 
convection, we expect that the meridional circulations should be weaker in 
lower mass stars and at faster rotation rates \citep{bro08,mat11,aug12}. 
In Babcock-Leighton flux-transport models, this should lead to long 
activity cycles \citep{jou10}. To test this assertion, we used the MESA 
code \citep{pax11} to generate a stellar structure model for 
$\epsilon$~Eri assuming a mass of 0.85~$M_\odot$ and an age of 0.8~Gyr. 
The radius, luminosity and $T_{\rm eff}$ of this model agree with the 
interferometric observations \citep{ba12}, and the convective velocities 
$v_c$ are roughly half as fast as in the solar convection zone. From 
angular momentum transport arguments ($v_m \sim v_c^2/\Omega$), this 
suggests that the meridional flow speed $v_m$ might be as small as 10\% of 
the solar value (or about 2--3 m\,s$^{-1}$ at the photosphere), making 
$\epsilon$~Eri a challenging case for flux-transport dynamos. Recently, 
3-D simulations of convectively driven dynamos in rapidly rotating stars 
have achieved large-scale organization and cyclic behavior in stellar 
convection zones \citep{bro11}. These dynamos do not rely on the slow 
meridional circulations, and can exhibit short cycles even in rapidly 
rotating lower mass stars, with the cycle period determined by the 
rotation rate and the convective properties \citep{nel13}. Of relevance to 
$\epsilon$~Eri, these convection zone dynamos show both short-period and 
long-period variations in the global-scale magnetism.

\section{DISCUSSION}\label{SEC4}

The long-term behavior of the magnetic activity cycles observed in 
$\epsilon$~Eri qualitatively resembles the interaction of the 11~year 
solar cycle with the quasi-biennial ($\sim$2~year) variations analyzed by 
\citeauthor{fle10}~(\citeyear{fle10}, their Fig.~1). The amplitude of the 
shorter (2.95~year) cycle in $\epsilon$~Eri appears to be modulated by the 
longer (12.7~year) cycle. During the broad minimum of the long cycle in 
1985--1992, there is no evidence of the short-period variations. 
\citeauthor{fle10}\ documented similar behavior in the Sun from 
helioseismic observations, with the quasi-biennial variations almost 
disappearing during the minimum of the 11~year solar cycle but gradually 
returning during the rise to solar maximum. They attributed this behavior 
to buoyant magnetic flux, generated near the tachocline during periods of 
high activity in the 11~year cycle, rising through the outer convection 
zone and episodically pumping up the amplitude of the quasi-biennial 
cycle. Support for this interpretation has emerged recently from efforts 
to localize the source of the quasi-biennial variations, placing them 
firmly in the near-surface regions of the Sun \citep{bro12}.

If we assume that the 2.95~year/12.7~year cycles in $\epsilon$~Eri are 
analogous to the 2~year/11~year cycles in the Sun, then the localization 
of the two signals has interesting consequences for the identification of 
the dynamos that are responsible for the active and inactive sequences as 
proposed by \cite{bv07}. She suggested that ``differential rotation near 
the surface mainly feeds [the A-sequence] dynamos'', while ``interface 
dynamos in the stars with deep [outer convection zones] are the important 
ones for the I-sequence stars''. This is precisely the opposite 
identification as that suggested by the helioseismic observations, which 
support a short cycle on the I-sequence localized in the near-surface 
regions and a long cycle on the A-sequence attributed to an interface 
dynamo at the tachocline. On the other hand, the Sun appears to be an 
outlier when compared to the A and I sequences established by observations 
of other stars. With the solar-like magnetic cycles observed in 
$\epsilon$~Eri, we are now in a better position to evaluate the specific 
property of the Sun that might make it peculiar in the context of other 
stars. Despite the fact that the Sun rotates less than half as fast as 
$\epsilon$~Eri, the stars each appear to have two interacting dynamos that 
operate on very similar timescales. This leads us to speculate that the 
rotational history of the Sun may be what makes it an outlier in the 
analysis of \cite{bv07}.

A complementary indication of the rotational history of a star can be 
found in its Li abundance. A stronger than usual rotational shear at the 
base of the outer convection zone can induce additional mixing below the 
tachocline where the temperature is sufficient to destroy Li. The solar Li 
abundance is anomalously low compared to some well-characterized solar 
twins such as 18~Sco \citep{mr07,baz11}, suggesting that the Sun may have 
been subjected to some additional mixing during its evolution. One way to 
explain a non-standard rotational history for the Sun is to blame Jupiter. 
\cite{bou08} suggested that a necessary condition for the formation of 
Jovian-mass planets is a long-lived proto-planetary disk, which then has 
sufficient time to interact with the stellar convection zone and induce a 
strong rotational shear between the radiative zone and the surface layers. 
This model predicts enhanced Li depletion among stars like the Sun and 
$\epsilon$~Eri which have Jovian-mass planets, compared to stars like 
18~Sco which do not. The Li abundance of $\epsilon$~Eri ($\log 
\epsilon$(Li)=$0.36\pm0.07$) is a factor of five below the solar value 
\citep[$\log \epsilon$(Li)=$1.05\pm0.10$;][]{asp09}, but it falls on the 
upper end of the distribution for stars with similar $T_{\rm eff}$ 
\citep{gon10}. Thus, if a Jovian-mass companion is responsible for the 
relatively low solar Li abundance, the underlying mechanism must not be 
effective in all cases.

Additional constraints on the interior structure and dynamics of 
$\epsilon$~Eri could come from future asteroseismic observations by the 
SONG network \citep{gru08}. Even without a dedicated network of 
telescopes, ground-based RV observations have already revealed solar-like 
oscillations in $\alpha$~Cen~B \citep{kje05}, which is only slightly more 
luminous than $\epsilon$~Eri. If solar-like oscillation amplitudes scale 
like $[L/M]^s$ \citep{kb95}, then the signal in $\epsilon$~Eri should be 
between 64\% \citep[s=1.5;][]{hou99} and 81\% \citep[s=0.7;][]{sam07} as 
strong as that seen in $\alpha$~Cen~B. Assuming comparable mode lifetimes 
for the two stars and using a more recent scaling relation that includes a 
dependence on $T_{\rm eff}$ \citep{kb11}, the estimate rises to 85\%. 
However, the relatively strong magnetic activity of $\epsilon$~Eri may 
suppress the oscillation amplitudes \citep{cha11}. Nearly uninterrupted RV 
measurements from SONG that span several months are expected to have the 
precision necessary to measure the depth of the surface convection zone 
\citep{vce06} and to reveal possible signatures of strong radial 
differential rotation \citep{gk93}. When combined with constraints on 
surface differential rotation from {\it MOST} \citep{cro06} and our 
detailed characterization of the interacting magnetic cycles, 
$\epsilon$~Eri may represent the best opportunity beyond the Sun to test 
stellar dynamo theory.


\acknowledgments We would like to thank David Gray, Steve Saar, Artie 
Hatzes, Fritz Benedict, and Barbara McArthur for helpful exchanges, as 
well as Phil Judge, Michael Kn\"olker, and Matthias Rempel for their 
contributions to the southern HK project. We are grateful to Fred Walter 
for scheduling our SMARTS program, and Manuel Hernandez, Jose Velasquez 
and Rodrigo Hernandez for conducting the observations at CTIO. The 
southern HK project has been supported under NOAO long-term programs 
2008B-0039 and 2011B-0001 with additional time from SMARTS partner 
institutions. CASLEO is operated under agreement between the Consejo 
Nacional de Investigaciones Cient{\'\i}ficas y T{\'e}cnicas de la 
Rep{\'u}blica Argentina and the National Universities of La Plata, 
C{\'o}rdoba and San Juan. B.P.B.\ is supported by NSF Astronomy and 
Astrophysics Postdoctoral Fellowship AST~09-02004. T.S.M.\ would like to 
thank the High Altitude Observatory for the severance that supported this 
work.






  \tablewidth{0pt}
  \tabletypesize{\footnotesize}
  \tablecaption{Journal of SMARTS observations for $\epsilon$~Eri.\label{tab1}}
  \begin{deluxetable*}{lcccc|lcccc}
  \tablehead{\colhead{DATE}&\colhead{UT}&\colhead{BJD\,(2450000+)}&\colhead{$S_{\rm MWO}$}&\colhead{$\sigma_S$}&
  \colhead{DATE}&\colhead{UT}&\colhead{BJD\,(2450000+)}&\colhead{$S_{\rm MWO}$}&\colhead{$\sigma_S$}}
  \startdata
  2007~Aug~22 & 08:37:02 & 4334.86078 & 0.5213 & 0.0023 & 2009~Nov~8  & 02:33:33 & 5143.61246 & 0.4971 & 0.0024 \\
  2007~Aug~22 & 08:37:56 & 4334.86141 & 0.5097 & 0.0022 & 2009~Nov~27 & 02:31:27 & 5162.61077 & 0.5497 & 0.0024 \\
  2007~Sep~17 & 07:11:59 & 4360.80377 & 0.5517 & 0.0027 & 2009~Nov~27 & 02:32:41 & 5162.61163 & 0.5592 & 0.0025 \\
  2007~Sep~17 & 07:13:03 & 4360.80451 & 0.5474 & 0.0027 & 2009~Dec~18 & 04:42:19 & 5183.70078 & 0.5807 & 0.0033 \\
  2007~Oct~8  & 07:51:16 & 4381.83230 & 0.5354 & 0.0021 & 2009~Dec~18 & 04:43:33 & 5183.70163 & 0.5944 & 0.0032 \\
  2007~Oct~8  & 07:53:28 & 4381.83383 & 0.5498 & 0.0020 & 2009~Dec~24 & 04:33:04 & 5189.69400 & 0.5532 & 0.0032 \\
  2007~Oct~25 & 03:50:35 & 4398.66577 & 0.5421 & 0.0017 & 2009~Dec~24 & 04:34:18 & 5189.69486 & 0.5559 & 0.0031 \\
  2007~Oct~25 & 03:52:49 & 4398.66732 & 0.5299 & 0.0020 & 2009~Dec~24 & 04:36:07 & 5189.69612 & 0.5502 & 0.0038 \\
  2007~Dec~12 & 02:25:59 & 4446.60645 & 0.5602 & 0.0015 & 2010~Jan~9  & 03:43:50 & 5205.65868 & 0.6342 & 0.0028 \\
  2007~Dec~12 & 02:28:13 & 4446.60800 & 0.5743 & 0.0016 & 2010~Jan~9  & 03:45:04 & 5205.65954 & 0.6392 & 0.0028 \\
  2007~Dec~20 & 06:08:27 & 4454.76051 & 0.5509 & 0.0017 & 2010~Feb~26 & 01:29:54 & 5253.56154 & 0.5974 & 0.0076 \\
  2007~Dec~20 & 06:10:41 & 4454.76206 & 0.5383 & 0.0020 & 2010~Feb~26 & 01:31:08 & 5253.56240 & 0.6209 & 0.0075 \\
  2008~Jan~3  & 02:01:24 & 4468.58805 & 0.5740 & 0.0016 & 2010~Mar~21 & 23:47:25 & 5277.48855 & 0.6373 & 0.0029 \\
  2008~Jan~3  & 02:03:18 & 4468.58937 & 0.5675 & 0.0015 & 2010~Mar~21 & 23:48:39 & 5277.48941 & 0.6298 & 0.0030 \\
  2008~Feb~29 & 01:06:20 & 4525.54498 & 0.5170 & 0.0024 & 2010~Mar~21 & 23:50:03 & 5277.49038 & 0.6285 & 0.0030 \\
  2008~Feb~29 & 01:07:34 & 4525.54583 & 0.5054 & 0.0024 & 2010~Aug~9  & 10:05:13 & 5417.92090 & 0.5800 & 0.0028 \\
  2008~Mar~17 & 00:01:35 & 4542.49871 & 0.4615 & 0.0021 & 2010~Aug~9  & 10:06:27 & 5417.92176 & 0.5806 & 0.0026 \\
  2008~Mar~17 & 00:02:49 & 4542.49956 & 0.4649 & 0.0024 & 2010~Aug~26 & 07:04:31 & 5434.79687 & 0.5639 & 0.0058 \\
  2008~Jul~9  & 09:30:20 & 4656.89418 & 0.4698 & 0.0038 & 2010~Aug~26 & 07:05:46 & 5434.79774 & 0.5539 & 0.0059 \\
  2008~Jul~9  & 09:31:34 & 4656.89503 & 0.4657 & 0.0038 & 2010~Aug~26 & 07:07:15 & 5434.79877 & 0.5675 & 0.0054 \\
  2008~Jul~27 & 09:36:41 & 4674.90003 & 0.5161 & 0.0027 & 2010~Sep~16 & 06:19:23 & 5455.76716 & 0.5674 & 0.0050 \\
  2008~Jul~27 & 09:37:55 & 4674.90088 & 0.5127 & 0.0027 & 2010~Sep~16 & 06:20:37 & 5455.76802 & 0.5662 & 0.0055 \\
  2008~Jul~27 & 09:39:09 & 4674.90174 & 0.5144 & 0.0027 & 2010~Sep~21 & 07:45:46 & 5460.82750 & 0.5768 & 0.0054 \\
  2008~Aug~8  & 08:14:46 & 4686.84416 & 0.4753 & 0.0049 & 2010~Sep~21 & 07:47:00 & 5460.82836 & 0.5676 & 0.0053 \\
  2008~Aug~8  & 08:16:00 & 4686.84502 & 0.4911 & 0.0064 & 2010~Oct~15 & 07:07:37 & 5484.80228 & 0.5304 & 0.0024 \\
  2008~Aug~18 & 10:05:14 & 4696.92175 & 0.4422 & 0.0018 & 2010~Oct~20 & 03:44:13 & 5489.66119 & 0.5487 & 0.0029 \\
  2008~Aug~18 & 10:06:28 & 4696.92261 & 0.4399 & 0.0017 & 2010~Oct~20 & 03:45:27 & 5489.66205 & 0.5353 & 0.0028 \\
  2008~Sep~7  & 05:48:54 & 4716.74537 & 0.4525 & 0.0060 & 2010~Nov~26 & 04:34:25 & 5526.69619 & 0.6010 & 0.0029 \\
  2008~Sep~7  & 05:50:08 & 4716.74623 & 0.4688 & 0.0029 & 2010~Nov~26 & 04:35:39 & 5526.69704 & 0.5999 & 0.0030 \\
  2008~Sep~15 & 06:25:08 & 4724.77113 & 0.4634 & 0.0020 & 2010~Dec~13 & 05:41:53 & 5543.74241 & 0.4972 & 0.0031 \\
  2008~Sep~15 & 06:26:22 & 4724.77199 & 0.4604 & 0.0020 & 2010~Dec~13 & 05:43:07 & 5543.74326 & 0.4843 & 0.0025 \\
  2008~Sep~25 & 05:08:51 & 4734.71882 & 0.4396 & 0.0018 & 2010~Dec~31 & 03:01:03 & 5561.62965 & 0.5582 & 0.0099 \\
  2008~Sep~25 & 05:10:05 & 4734.71968 & 0.4430 & 0.0018 & 2010~Dec~31 & 03:02:17 & 5561.63050 & 0.5112 & 0.0111 \\
  2008~Oct~4  & 05:39:08 & 4743.74037 & 0.4288 & 0.0026 & 2011~Jan~8  & 03:28:43 & 5569.64827 & 0.5704 & 0.0030 \\
  2008~Oct~4  & 05:40:22 & 4743.74122 & 0.4363 & 0.0026 & 2011~Jan~8  & 03:29:57 & 5569.64913 & 0.5492 & 0.0028 \\
  2008~Oct~12 & 05:37:24 & 4751.73953 & 0.4681 & 0.0050 & 2011~Jan~30 & 03:25:41 & 5591.64433 & 0.5673 & 0.0067 \\
  2008~Oct~12 & 05:38:37 & 4751.74038 & 0.4657 & 0.0052 & 2011~Jan~30 & 03:26:55 & 5591.64519 & 0.5520 & 0.0058 \\
  2008~Oct~25 & 06:23:37 & 4764.77205 & 0.5247 & 0.0159 & 2011~Feb~26 & 00:42:39 & 5618.52875 & 0.5410 & 0.0068 \\
  2008~Oct~25 & 06:24:51 & 4764.77291 & 0.4274 & 0.0085 & 2011~Feb~26 & 00:43:53 & 5618.52960 & 0.5237 & 0.0058 \\
  2008~Nov~2  & 07:06:32 & 4772.80199 & 0.4799 & 0.0031 & 2011~Mar~9  & 00:41:04 & 5629.52675 & 0.5690 & 0.0033 \\
  2008~Nov~2  & 07:07:46 & 4772.80285 & 0.5108 & 0.0033 & 2011~Mar~9  & 00:42:18 & 5629.52761 & 0.5662 & 0.0028 \\
  2008~Nov~10 & 05:37:51 & 4780.74045 & 0.4464 & 0.0020 & 2011~Jul~20 & 09:47:19 & 5762.90675 & 0.4732 & 0.0056 \\
  2008~Nov~10 & 05:39:05 & 4780.74130 & 0.4558 & 0.0020 & 2011~Jul~20 & 09:48:33 & 5762.90761 & 0.4827 & 0.0050 \\
  2008~Nov~26 & 06:30:29 & 4796.77678 & 0.4713 & 0.0027 & 2011~Sep~1  & 08:21:44 & 5805.85096 & 0.4406 & 0.0038 \\
  2008~Nov~26 & 06:31:42 & 4796.77763 & 0.4737 & 0.0027 & 2011~Sep~1  & 08:22:58 & 5805.85182 & 0.4442 & 0.0037 \\
  2008~Dec~2  & 04:34:49 & 4802.69628 & 0.4392 & 0.0021 & 2011~Sep~18 & 07:54:44 & 5822.83350 & 0.4520 & 0.0030 \\
  2008~Dec~2  & 04:36:03 & 4802.69714 & 0.4329 & 0.0021 & 2011~Sep~18 & 07:55:58 & 5822.83436 & 0.4542 & 0.0033 \\
  2008~Dec~10 & 01:06:35 & 4810.55136 & 0.4552 & 0.0046 & 2011~Oct~22 & 07:49:41 & 5856.83171 & 0.4010 & 0.0066 \\
  2008~Dec~10 & 01:07:49 & 4810.55222 & 0.4456 & 0.0071 & 2011~Oct~22 & 07:50:55 & 5856.83257 & 0.4276 & 0.0061 \\
  2009~Jan~13 & 03:18:41 & 4844.64089 & 0.4327 & 0.0020 & 2011~Nov~14 & 06:25:34 & 5879.77356 & 0.4517 & 0.0024 \\
  2009~Jan~13 & 03:19:55 & 4844.64175 & 0.4310 & 0.0019 & 2011~Nov~14 & 06:26:48 & 5879.77441 & 0.4486 & 0.0024 \\
  2009~Jan~20 & 02:31:57 & 4851.60786 & 0.4137 & 0.0024 & 2011~Dec~4  & 03:03:34 & 5899.63286 & 0.4001 & 0.0020 \\
  2009~Jan~20 & 02:33:11 & 4851.60872 & 0.4178 & 0.0025 & 2011~Dec~5  & 02:45:36 & 5900.62034 & 0.4231 & 0.0021 \\
  2009~Jan~31 & 02:45:28 & 4862.61629 & 0.4327 & 0.0020 & 2011~Dec~5  & 02:46:50 & 5900.62120 & 0.4212 & 0.0021 \\
  2009~Jan~31 & 02:46:42 & 4862.61715 & 0.4254 & 0.0020 & 2011~Dec~14 & 03:04:15 & 5909.63290 & 0.3987 & 0.0020 \\
  2009~Feb~8  & 01:34:25 & 4870.56624 & 0.4655 & 0.0021 & 2011~Dec~14 & 03:05:29 & 5909.63376 & 0.3916 & 0.0019 \\
  2009~Feb~8  & 01:35:39 & 4870.56710 & 0.4519 & 0.0020 & 2011~Dec~23 & 02:35:50 & 5918.61267 & 0.4203 & 0.0039 \\
  2009~Mar~28 & 23:49:25 & 4919.48951 & 0.4767 & 0.0043 & 2011~Dec~23 & 02:37:04 & 5918.61353 & 0.4240 & 0.0049 \\
  2009~Mar~28 & 23:50:39 & 4919.49036 & 0.4586 & 0.0024 & 2012~Jan~14 & 04:11:35 & 5940.67758 & 0.4219 & 0.0021 \\
  2009~Jul~1  & 10:03:32 & 5013.91665 & 0.5200 & 0.0025 & 2012~Jan~14 & 04:12:49 & 5940.67844 & 0.4243 & 0.0021 \\
  2009~Jul~1  & 10:04:46 & 5013.91751 & 0.5155 & 0.0027 & 2012~Jan~28 & 02:34:39 & 5954.60909 & 0.4097 & 0.0032 \\
  2009~Aug~28 & 07:19:32 & 5071.80750 & 0.5136 & 0.0023 & 2012~Jan~28 & 02:35:53 & 5954.60994 & 0.4111 & 0.0031 \\
  2009~Aug~28 & 07:20:46 & 5071.80835 & 0.5121 & 0.0027 & 2012~Jul~9  & 10:28:07 & 6117.93430 & 0.4836 & 0.0023 \\
  2009~Sep~27 & 07:00:06 & 5101.79619 & 0.5792 & 0.0027 & 2012~Jul~9  & 10:29:22 & 6117.93517 & 0.4877 & 0.0023 \\
  2009~Sep~27 & 07:01:22 & 5101.79707 & 0.5707 & 0.0028 & 2012~Aug~9  & 09:18:52 & 6148.88876 & 0.5407 & 0.0033 \\
  2009~Oct~17 & 05:08:16 & 5121.71948 & 0.4986 & 0.0030 & 2012~Aug~9  & 09:20:07 & 6148.88963 & 0.5278 & 0.0032 \\
  2009~Oct~17 & 05:09:30 & 5121.72034 & 0.4916 & 0.0030 & 2012~Sep~3  & 08:59:35 & 6173.87748 & 0.5024 & 0.0031 \\
  2009~Nov~3  & 05:31:51 & 5138.73625 & 0.5710 & 0.0036 & 2012~Sep~3  & 09:00:50 & 6173.87835 & 0.5111 & 0.0031 \\
  2009~Nov~3  & 05:33:05 & 5138.73710 & 0.5670 & 0.0032 & 2012~Oct~13 & 07:16:28 & 6213.80837 & 0.5753 & 0.0042 \\
  2009~Nov~3  & 05:34:28 & 5138.73806 & 0.5624 & 0.0036 & 2012~Oct~13 & 07:17:42 & 6213.80923 & 0.5738 & 0.0040 \\
  2009~Nov~8  & 02:32:20 & 5143.61161 & 0.5069 & 0.0024 & \nodata     & \nodata  & \nodata     &\nodata & \nodata 
  \enddata
  \end{deluxetable*}

\end{document}